# NEGATIVE PARTICLE PLANAR AND AXIAL CHANNELING AND CHANNELING COLLIMATION[†]

RICHARD A. CARRIGAN, JR.
*Fermi National Accelerator Laboratory[*], Batavia, IL 60510, USA*

While information exists on high energy negative particle channeling there has been little study of the challenges of negative particle bending and channeling collimation. Partly this is because negative dechanneling lengths are relatively much shorter. Electrons are not particularly useful for investigating negative particle channeling effects because their material interactions are dominated by channeling radiation. Another important factor is that the current central challenge in channeling collimation is the proton-proton Large Hadron Collider (LHC) where both beams are positive. On the other hand in the future the collimation question might reemerge for electron-positron or muon colliders. Dechanneling lengths increase at higher energies so that part of the negative particle experimental challenge diminishes. In the article different approaches to determining negative dechanneling lengths are reviewed. The more complicated case for axial channeling is also discussed. Muon channeling as a tool to investigate dechanneling is also discussed. While it is now possible to study muon channeling it will probably not illuminate the study of negative dechanneling.

## 1. The Potential for Tev-Scale Channeling

The potential for TeV-scale channeling is significant. The challenge of LHC collimation has been an important driver for new developments. The possible use of channeling collimation has only become more intriguing with recent work at CERN. The recently illuminated coherent processes volume reflection and volume capture are interesting from both a fundamental channeling perspective as well as for applications. While these ideas go back to the era of work by Vorobiev and Taratin[1] it is only in the last few years that compelling experimental information has emerged at RHIC[2], at Fermilab[3] and now in a spectacular series of experiments at CERN[4]. Experiments on collimation and volume effects have been carried out at both CERN and the Fermilab Tevatron in the very recent past. Theoretical studies of collimation as well as volume effects are underway at many places. An important breakthrough is the development and use of short crystals. Other questions and applications for TeV-range channeling have been investigated. Channeling could be used for extraction[5]. Negative particle channeling may be relevant to channeling collimation for muon machines and an ILC that will use electrons as well as positrons.

[†] Presented at *Channeling 2008*, Erice, Italy, 25 October – 1 November, 2008.
[*] Operated by the Fermi Research Alliance, LLC under Contract No. DE-AC02-07CH11359 with the United States Department of Energy.

Fermilab channeling collimation studies began in 2005. Actually the interest harks back to the work of Mokhov and others at the SSC in the nineties[6]. The trigger for studies at Fermilab was also partly the curious RHIC results which showed anti-collimation but also what has turned out to be volume reflection. Fermilab reproduced and extended the RHIC work in 2005 with an “O” shaped crystal. A strip crystal was tried next but there were difficulties with the goniometer and perhaps the crystal and other matters. The studies have returned to the old crystal with a new goniometer and instrumentation. The investigations are getting even more interesting results. T980 or CCE, a crystal collimation experiment with LARP backing (the U.S. LHC Accelerator Research Program) and groups from Fermilab, SLAC, CERN, BNL, INFN, IHEP, PNPI, JINR, RINP-BSU, Chicago, and others is now underway. It is proceeding in several stages over the 2008-2010 period. It is also linked with a complementary SPS program at CERN to employ artificially produced halo. These investigators may even study single particle behavior.

The following sections discuss dechanneling including dechanneling estimates, negative particle dechanneling, studying dechanneling, and a reprise on the possibilities for future negative particle dechanneling studies. The reminder of the article is devoted to sections on axial channeling including the possibility of axial bending, muon channeling, and a summary.

## 2. Dechanneling

Dechanneling theory had been fairly well developed by the time investigations of channeling in the multi-hundred GeV regime began at Fermilab in the mid seventies. Gemmell’s[7] review of the work of Lindhard[8] and others covers this early period in detail. In that period the channeling field mainly emphasized non-relativistic beams with energies in the MeV range used for nuclear and solid state physics applications. Two important topics were not included in the developments up to that point. These were the possibility of bent crystal channeling suggested by Tsyganov[9] while working on the first Fermilab channeling experiment and the subject of heavy negative particle channeling. Practically and also theoretically electron and heavy negative particle dechanneling are quite different subjects. In all of these developments channeling was considered to be the process that occurred in a crystal out to an angle of order the Lindhard critical angle. Volume reflection and volume capture, subjects closely related to relativistic channeling have also appeared since the Lindhard-Gemmell era.

Tsyganov's original discussion of bent crystal channeling in 1976 contains some discussion of the negative particle case. His second note has a more detailed treatment of planar bending. In these notes the critical radius, now often called the Tsyganov radius, is defined in terms of the electric field as a gradient of a crystal potential.

The advent of bent crystal channeling facilitated dechanneling studies for heavy particles in the multi-hundred GeV regime. Detailed experimental studies of bending and bending dechanneling were underway by the early eighties at CERN, Serpukhov, and Fermilab. Some of these studies are summarized by Carrigan[10] and in the book by Birykov, Chesnokov, and Kotov[11] (BCK).

The recent multi-hundred GeV channeling results from the CERN SPS experiments in H8 and H4 have produced results on negative particle deflection and dechanneling. The H8/H4 results are covered in the theses of Hasan[12] and Bolognini[13].

### 2.1. *Dechanneling Estimates*

A theoretical evaluation of negative particle dechanneling is needed to assay current and future channeling applications and experiments. Ideally this would be in terms of a diffusion treatment in the spirit of the diffusion calculation for electron dechanneling by Backe et al.[14] or Maisheev's recent attack on volume reflection[15]. A second approach would be to use a simulation program such as BINCOL[16] or CATCH[17] to map out a matrix of crystal bending radii and beam energy values. An alternative approach is to use a phenomenological model that compares the multiple scattering in a channel to the focusing effect of the lattice. Finally, as noted below, it is important that dechanneling models be able to accommodate bent crystals.

The dechanneling length reflects the diffusion out of a channel compared to the angular distribution inside the channel which is related in turn to the channeling potential well. The larger the multiple scattering in the channel, the shorter the dechanneling length. With this perspective Feldman and Appleton[18] developed a diffusion picture from Lindhard[8] and converted it to the phenomenological dechanneling length

$$\lambda_D = 1.62 \frac{\psi_{cp}^2}{\left\langle \Theta^2 \right\rangle_c} \tag{1}$$

where $\psi_{cp}$ is the planar critical angle and $<\theta^2>_c$ is the mean square multiple scattering angle in the channel.

BCK uses this to get

$$\lambda_D = \frac{256}{9\pi^2} \frac{pv}{\ln(2m_e c^2 \gamma / I) - 1} \frac{a_{TF} d_p}{Z_i r_e m_e c^2}. \quad (2)$$

where $Z_i$ is the incident particle charge, $d_p$ is the interplanar distance and $a_{TF}$ is the Thomas-Fermi screening length. At 1 TeV this leads to $\lambda_D$ = 510 mm for (110) planar proton channeling in Si. This dechanneling length roughly goes as E (pv). The logarithmic term actually changes by O(10%) over a decade so that a linear interpolation with energy is only an approximation. Note that this functional form would have to be modified for positrons in part because of relativistic kinematic considerations and in part because of electronic multiple scattering. If these factors are ignored straight application of the formula gives a planar dechanneling length of 400 microns at 855 MeV for planar positron channeling in Si.

### 2.2. *Negative Particle Dechanneling*

There has been little study of negative hadron channeling either theoretically or experimentally. Part of this has been due to the facts that negative particle dechanneling lengths have been expected to be short. Higher energy and shorter crystals can diminish this problem. For electrons channeling and channeling radiation are interwoven so that the impact of channeling radiation on applications needs to be better understood.

The possibility of negative particle deflection with bent crystals may have been discounted earlier because the applications that were considered required large discrete angular deflections. The situation is different for collimation where the important point is to give particles a kick, any kick, provided it is more than multiple scattering. For some applications multi-pass channeling and high energy also help.

As noted earlier, dechanneling estimates can be developed using a formal diffusion model, multiple scattering phenomenology, or computer simulations. Below multiple scattering phenomenology is considered first, then diffusion modeling.

#### 2.2.1. *Phenomenological model*

The dechanneling equation introduced above (Equation 1) can be used to infer negative particle dechanneling lengths. Two sets of factors are needed, the multiple scattering in the channel for positive and negative particles and the critical angles.

The fact that the ratio of the Lindhard angles for positive particle channeling to the negative case is roughly a factor of two has been recognized for many years. The left-hand panel of Figure 1 shows a schematic harmonic oscillator planar potential for positive particles. (A more sophisticated potential is chosen for most channeling theories.) The shaded areas indicate the regions occupied by the nuclear centers in the lattice. The half width is $u_T$, the rms thermal vibration amplitude. This is temperature dependent. For silicon under typical conditions it is 0.075 Å. In the right-hand panel the potential for a negative particle is inverted. Now the nuclear centers are in the middle of the channel so the multiple scattering will increase. Notice that the effective channel width is narrower for the negative case. Perhaps the first observation of this was in a study by Uggerhoj and Andersen in 1968[19] using implanted radioactive $^{64}$Cu atoms in a copper lattice. This radioactive source produces both positive and negative beta rays. The negative channeling features are roughly half the width of the positive ones. Note that these results are for MeV-scale particles. Almost thirty-five years later the Aarhus group published results for MeV-scale antiprotons[20]. While this measurement was not a side-by-side comparison, the antiproton distribution width shows the expected narrowing and agrees with theoretical estimates.

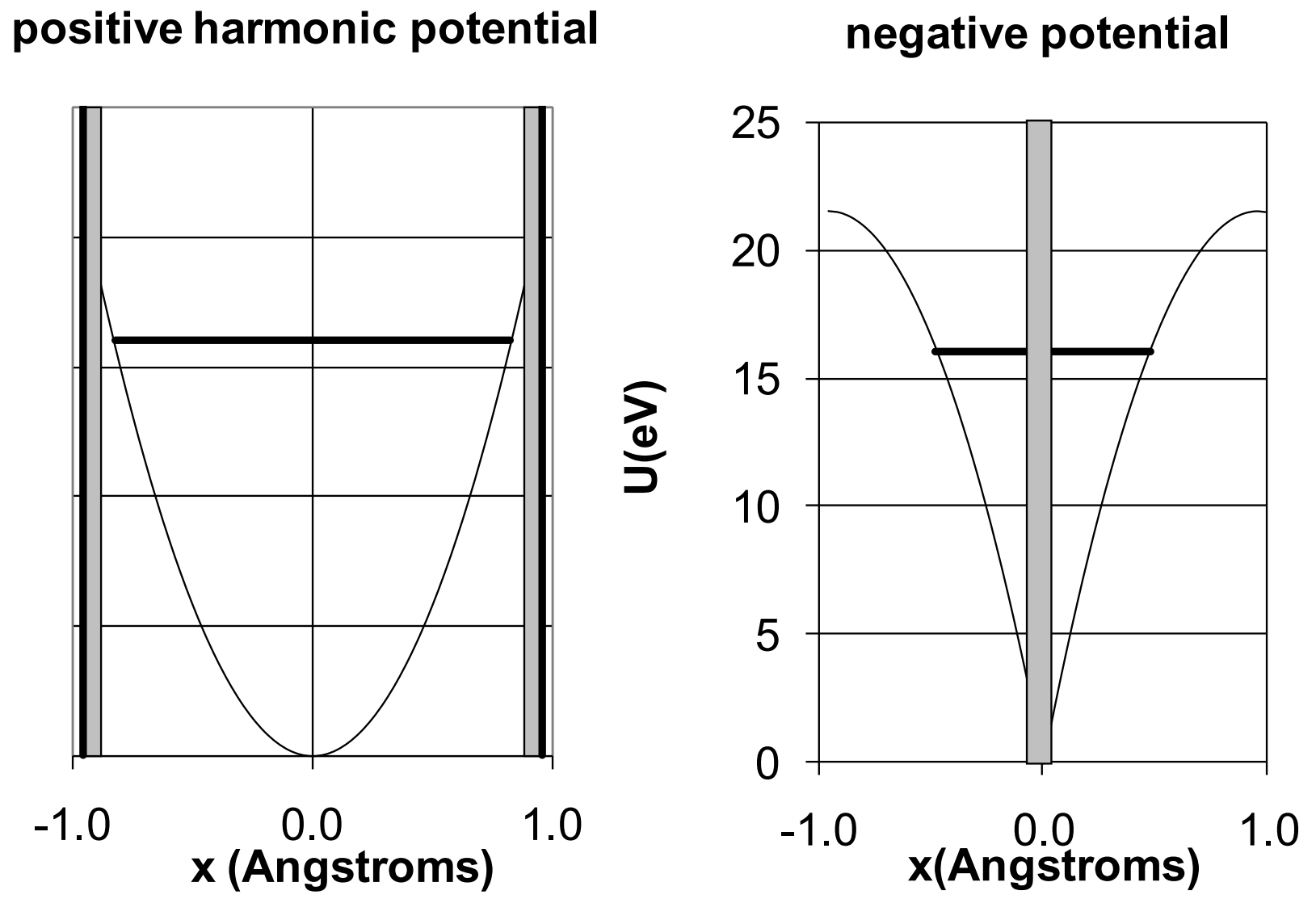

Figure 1. Schematic potential for Si(110) planar geometry. Left is for a positive particle, right for a negative particle.

Multiple scattering for positive particles is due only to electrons in the channel while the multiple scattering for negative particles must factor in the ions in the channel. For the moment we assume the nuclear density fills the channel. The ratio of the multiple scattering angles for the positive and negative cases is

$$\frac{\left.\frac{d\Omega^2}{dx}\right|_e}{\left.\frac{d\Omega^2}{dx}\right|_n} = \frac{\left.\frac{d\Omega^2}{dx}\right|_+}{\left.\frac{d\Omega^2}{dx}\right|_-} = \frac{(m_e/2M_iE)(4\pi e^4/m_e v^2)NZL_e}{(M_t/M_iE)(4\pi Z^2 e^4/M_t v^2)NL_n} = \frac{L_e}{2ZL_n} \Rightarrow \frac{X_0^n}{X_0^e} \quad (3)$$

where $L_i$ are the log terms in multiple scattering, Z is the target charge, and N is the number of atoms per unit volume. $L_n/L_e$ has been set to 2. The $X_0$ are the radiation lengths for the electron and ionic cases. Note that this formula is slightly confusing because the positive particle multiple scattering is due to the electrons in the channel.

This leads to a negative particle dechanneling length:

$$\lambda_- = \left(\frac{\psi_-}{\psi_+}\right)^2 \frac{\lambda_+}{2Z(L_n/L_e)}, \quad (4)$$

where $\lambda_+$ is given by, for example, equation (2). Neglecting details of the nuclear density profile Equation (4) gives for Si(110) at 1 TeV:

$$\lambda_- = \left(\frac{\psi_-}{\psi_+}\right)^2 \frac{\lambda_+}{2Z(L_n/L_e)} = \left(\frac{0.48\,A}{0.83\,A}\right)^2 \frac{51\,cm}{2*14*2} = 3\;mm$$

That is, with the naïve assumption that the ionic density covers the channel the negative dechanneling length at 1 TeV is a short 3 mm. For 855 MeV $\lambda_-$(filled) is 2.5 microns.

But the negative potential well is filled with ions only in the region $\pm u_T$. One could weight the earlier formula proportional to the space covered by thermal vibrations, that is:

$$Z_{eff}^- = \frac{2u_T Z + 2(d_p/4 - u_T)}{d_p/2}. \quad (5)$$

This gives a Z effective of about 3 rather than 14. This leads to $\lambda_-$(1 TeV) ~ 14 mm. For this situation the experimental ratio is $\lambda_+/\lambda_-$ ~ 510 mm/14 mm or 36. For 855 MeV $\lambda_-$(not filled) is 12 microns. Of course a typical channeled particle spends more time in the center of the channel. However the ion density in the center could have a higher impact than this weighting gives. A full calculation of the process might give a length somewhere in between.

### 2.2.2. *Diffusion model*

As noted earlier, Backe et al.[14] have recently used a Fokker-Planck diffusion model to obtain a value for the electron dechanneling length at 855 MeV. Their treatment is based on Kumakhov-Komarov[21], Beloshitsky-Trikalinos[22], and the Ohtsuki-Kitagawa[23] approximation.

In assaying a dechanneling theory one important consideration is establishing a link to a practical dechanneling length that includes the ramifications of microscopic phase space over the channel. A second consideration is the nature of the physical process used to monitor the yields (such as channeling radiation or bent crystal channeling). There are many caveats and reflections on how Backe, et al. arrive at a solution. They calculate a dechanneling length of 18.4 microns for (110) silicon at 855 MeV (deep in the crystal). Parenthetically, the measured length for the so-called "high energy case" is 32 ± 4 microns. Based on their 855 and 1200 MeV points this goes as 21.3 microns/GeV. A linear extrapolation approximation gives 21 mm for the negative particle dechanneling length at 1 TeV. Interestingly Backe et al. did not see linear scaling for the beam energy region around 150 MeV.

It could be quite useful to analyze the negative particle hadronic dechanneling case employing the same approach used by Backe, et al. for the $e^+/e^-$ analysis. It would appear relatively straightforward to adapt the calculation to heavy particles, both positive and negative. Some of problems associated with the radiative processes would no longer be present. If possible, it would be desirable to carry out such a calculation so it shows beam energy explicitly. That is to say, the computed dechanneling length should be given as a function of beam energy either via a functional analytic form or a table of dechanneling length as a function of energy.

## 2.3. ***Studying Dechanneling***

In the multi-hundred GeV regime most dechanneling measurements and estimates employ bent crystal channeling. The dechanneling length is found either by determining the number of particles that complete the bend or by examining the distribution of the particles lost in going around the bend. Both approaches are obviously based on approximations.

For curved crystals the critical bending radius is given by Tsyganov's formula:

$$R_T \approx \frac{pv}{\pi N d_p Z e^2} = \frac{pv(GeV)}{5.7 GeV/cm[Si(110)]}. \tag{6}$$

At 1 TeV in silicon the Tsyganov radius is approximately 175 cm.

The dechanneling length in a bent crystal with radius of curvature R is (BCK):

$$\lambda_D(pv, R) = \lambda_{D,0}\left(1 - \frac{R_T}{R}\right)^2 . \qquad (7)$$

For the negative case the effective $d_p$ would be smaller and the Tsyganov radius would be larger. The straight crystal dechanneling length is given by (2) or (4) above as appropriate. If R is much greater than $R_T$ the bending dechanneling length is approximately equal to the straight crystal dechanneling length and a bent crystal can be used to evaluate it. From the earlier discussion this length is shorter for negative particles so a short crystal must be used.

### 2.3.1. *Experimental hadronic planar dechanneling measurements based on bending dechanneling:*

Data from the recent CERN H8 and H4 bent crystal channeling studies can be used to estimate dechanneling lengths. Although the negative dechanneling lengths are short the crystals used in the H8 and H4 measurements have been on the order of 1 mm long. (Information on these measurements is contained in the theses of Hasan[12] and Bolognini[13].)

The CERN H8/H4 experiment can be used to make a rough estimate for the positive dechanneling length for the (110) planar case with 400 GeV protons and the 3 mm long ST4 crystal. For this case Figure 5 in Scandale et al. [24] can be used to extract the dechanneled fraction and fraction of the beam that is channeled. That estimate is then multiplied by two to reflect the more reasonable cut used for Figure 6 in the article. This leads to an estimate for the dechanneling length of $\lambda_d$ = 200 mm. A linear extrapolation with energy gives a dechanneling length of 500 mm at 1 TeV. Parenthetically, Equation (2) based on BCK gives 510 mm at 1 TeV. Simulations in BCK section 3.6.3 extrapolated to 1 TeV give 560 mm for (110) and other measurements from 70 GeV extrapolated to 1 TeV give 600 mm (BCK p. 103). Thus for positive particle dechanneling the recent H8 work is consistent with earlier dechanneling measurements.

### 2.3.2. *The ratio of negative to positive dechanneling lengths*

Data from Bolognini's thesis (Figure 4.21a) for the 150 GeV negative particle case (mainly pions) suggests $\lambda \sim O(3$ mm) for the (111) plane in a quasi mosaic crystal and also for a (110) strip crystal. A linear extrapolation to 1 TeV yields a dechanneling length of 20 mm.

The experimental ratio for the positive and negative particle dechanneling lengths taken from the H4/H8 experiment for heavy particles is

$\lambda_+/\lambda_- \sim$ 200 mm/((400/150)*3 mm) or 25.

In summary, some of the determinations of the negative particle dechanneling lengths for the (110) case are given in Table 1 below.

Table 1: Negative particle dechanneling lengths

| | lamda (mm) | Comments |
|---|---|---|
| Backe-Lauth | .0184 | theory-electrons/positrons (0.855 GeV) |
| H4@CERN SPS | 3 | 150 GeV experiment |
| Linear extrapolation to 1 TeV | | |
| Backe-Lauth | 21 | theory-electrons/positrons |
| BCK (Carrigan) | 14 | theory/ansatz-hadrons-ions fill to $\pm u_T$ |
| H4@CERN SP | 20 | experiment-hadrons (150 GeV) |

The ansatz for the negative particle length based on BCK is low but not so far from Backe-Lauth and the H4 data. Obviously the Backe-Lauth case done for leptons examines a somewhat different process than the hadronic dechanneling case.

### 2.4. *Future negative particle dechanneling studies*

More information is needed about negative particle channeling and negative bending! Several distant but still possible situations exist where negative particle collimation might be interesting. For example, could one channel and collimate antiprotons at the Tevatron? In an experiment like TOTEM at the LHC could one deflect negative particles such as pions or kaons? Could one collimate both $e^-$ and $e^+$ at an ILC?

As noted earlier the CERN H4 negative particle beam has already provided some information on negative particle dechanneling. It is important to extract as much information as possible from the existing data set. It will be very interesting to hear the results from that experiment. Negative particle beams at Serpukhov and Fermilab run at lower energies than H4 so that they would be less likely to provide useful information.

The highest energy negative particle beam available in the world is the circulating antiproton beam at the Fermilab Tevatron. Because the negative particle dechanneling length is short "highest energy" is a useful feature for studying this process. Even "long" 5 mm crystals would probably still show bending channeling. D. Still at Fermilab has looked at the possibility of studying dechanneling with antiprotons circulating in the Tevatron. At present the goniometer and detectors used for the collimation studies are on the proton beam side. This aspect could be overcome by redoing the so-called proton and antiproton helical orbits in the vicinity of E0 where the equipment is located. Alternatively the proton beam could be removed from the accelerator at the end of a store and studies carried out using the antiprotons. Detectors and collimation devices would be required in the antiproton downstream direction. Either rearranging the Tevatron beams at E0 or removing the proton beam requires special efforts. Some clever ideas to lessen the impacts might be uncovered.

In general experimental information on dechanneling of electrons is less useful for hadronic programs. This is because radiative processes are mixed in with dechanneling effects. On the other hand the information is potentially important for the design of devices based on crystalline undulators.

Note that at this time there is still little experimental information on negative particle dechanneling in a straight crystal.

## 3. THE AXIAL CASE

As noted earlier critical angles for axial channeling are typically several times larger than planar critical angles. As a result, the angular acceptance for channeling is larger. Equation (1) above indicates that the dechanneling lengths would also be longer. On the other hand axial channeling is a more complicated process. It is also more complicated to exploit for applications because it requires alignment for two different angular degrees of freedom. In axial channeling particles can move in the minimum of the potential well between several strings or, at larger angles to the axis, meander through the lattice.

Investigations of high energy axial behavior in straight crystals were reported in the early eighties by an Aarhus-CERN group[16] and the channeling collaboration at Fermilab[25]. Beam flux distributions for both positive and negative particles aligned with an axis showed a peak around the axis. The outgoing distribution formed a doughnut for beams slightly away from the axis. Beyond the Lindhard critical angle the doughnut was only partially filled. Negative particle effects extended to somewhat larger angles. Both groups interpreted this axial behavior as a statistical equalization process in transverse momentum as outlined by Lindhard[8] in the sixties. A similar diffusion in transverse energy was also observed. For the positive case particles leaked out into planes. No planar effects were seen for negative particles (see the review by Akhiezer et al.[26]). It is because of the possibility of both diffusion out into a random direction and also leakage into a plane that axial dechanneling is a more complicated process.

Tsyganov's treatment of bent crystal channeling was developed in 1976. Tsyganov's first paper does mention the possibility of axial bending. Parenthetically Schiott[27] notes that J. U. Andersen had considered axial bending due to doughnuts in 1979.

In the 1980s the Aarhus group[28] also carried out an axial bending experiment for both positive and negative particles at 12 GeV/c. They note that the mean free path for transverse momentum equalization, $\lambda_t$, is:

$$\lambda_t \cong \frac{4\psi}{\pi^2 Nda_{TF}\psi_1^2}, \quad (8)$$

where $\psi_1$ is the axial critical angle and d is the interatomic spacing along the string. Note that this formula is linear in $\psi$, the angle of the doughnut to the axis. For $\psi$ values approaching the critical angle $\lambda_t$ is about 10 microns at 12 GeV/c. The Aarhus group used the diffusion equation in transverse energy to derive an axial dechanneling length in a crystal bent with radius of curvature R:

$$l_d \cong \frac{R^2\psi_1^2}{2\lambda_t}. \quad (9)$$

This can be used to develop a necessary condition for axial channeling in a bent crystal of

$$\frac{R\psi_1}{\lambda_t} > 1. \quad (10)$$

While this condition was fulfilled at 12 GeV/c the axial dechanneling length was much shorter that the crystal lengths of 20-30 mm. In this experiment where the energy was small and the effective bend radius was closer to the Tsyganov radius the amount of deflection was small, on the order of the critical angle. Schiott reproduced the experimental results with the BINCOL simulation routine.

The Aarhus-CERN axial bending investigations were extended to higher energies[29] in the mid nineties when they carried out a study using 200 GeV/c negative pions. The conditions of the experiment were such that (10) was not satisfied so axial bending through the full bending angle was not expected. From the Aarhus-CERN studies it is possible to posit an upper limit on the bending dechanneling length of O(2 mm). Doing this from the available information in a realistic way is difficult.

In the nineties Shul'ga, Greenenko, and colleagues undertook a comprehensive theoretical investigation of axial bending. In these articles they used "channeling" for what had earlier been called hyperchanneling and "above-barrier" to refer to the channeling region in the vicinity of the critical angle. A review article by Akhiezer et al.[26] and an article by Greenenko and Shul'ga[30] herald the later work.

The key article[31] in the series develops a number of formulas and compares the functional forms to the Aarhus[28] treatment. This development is further amplified in Greenenko and Shul'ga[32]. The length for beam transverse momentum equalization as defined in Greenenko and Shul'ga is:

$$l_t \approx \frac{1}{\psi_1 Nda_{TF}}. \quad (11)$$

Outside of constants $l_t$ differs from $\lambda_t$ by $\psi/\psi_1$. This is a non-trivial difference. Greenenko and Shul'ga[32] develop a condition "for realization of the bending effect" in their treatment:

$$\alpha = \frac{L}{R\psi_1}\frac{l_t}{R\psi_1} < 1 \tag{12}$$

where L is the crystal length. This includes the additional term $L/R\boldsymbol{\psi_1}$ relative to the Bak et al. expression given above in (10). Greenenko and Shul'ga note that according to their constraint (12) the 1996 Aarhus-CERN negative axial bending experiment[29] was not in the correct regime to observe axial deflection.

There are several other articles in the Shul'ga and Greenenko series. Greenenko and Shul'ga[33] is directed at larger crystal thicknesses. It contains some details on their model including the recursion relation. It has simulations for the <111> axis where the energy is +/-100 GeV, R is set to 100 m, and the crystal lengths were one and three cm. The negative motion case is diffuse while the positive case shows leakage out into the planes. Elsewhere they have a low statistics <111> axial simulation for +/-800 GeV, R = 3*10^4 cm, and a crystal length of three cm. Greenenko and Shul'ga[34] extends these concepts for a 450 GeV study. Perhaps for the first time the article shows the development of doughnuts as a function of crystal length.

The work of Greenenko and Shul'ga[35] does contain <110> axial simulations for conditions that agree with the Aarhus-CERN results of Baurichter et al. The same approach also yields agreement with the recent CERN H8[36] results for a 2 mm thick silicon crystal bent through 50 microradians in a 400 GeV/c proton beam aligned with the <111> axis. The data in the H8 study clearly demonstrates axial bending for positive particles. According to the article the conditions of the experiment were such that the Shul'ga-Greenenko condition was satisfied. This demonstration of axial bending shows that it may be possible to exploit short crystals for more efficient collimation.

For future collimation activities it would be useful to extend the treatment of bent crystal axial dechanneling. In particular it would help to have practical dechanneling measures that take into account conventional bending dechanneling, leakage to planes, and constraints on doughnut effects. It should be noted that the CERN RD22 H8/H4 collaboration has already established some useful measures for deflection efficiency[36].

## 4. Muon Channeling

Several years ago I wondered if studying muon channeling might answer questions that could be interesting for negative particle channeling[37]. The argument went that negative particles would encounter more nuclei moving through a channel so that effectively the nuclear interaction length would be shorter. Doing separate studies of muons and strongly interacting negative particles would illuminate the situation. In fact while the nuclear density in the channel might be five to ten higher than say a random direction, the nuclear collision length for negative particles is still 30 cm in Si for a random direction while the negative particle dechanneling length is 1-100 mm depending on energy. Except for energies beyond 10 TeV negative particles dechannel before they interact! The bremsstrahlung cross section for muons is down by a factor of 40,000 relative to electrons. The critical energy for muons in silicon is 470 GeV. "Channeling radiation" for muons is probably a relatively insignificant problem even well into the trans TeV regime.

While muon channeling appears not to be a significant study tool for dechanneling the possibility of higher energy muon colliders is now discussed for the far future. There collimation may once again become a challenge and channeling might have something to offer. Muon channeling studies are difficult but the CERN experiment in H8 already has the capacity to tag muons with the DEVA energy loss calorimeter[38].

## 5. Summary

In summary although negative particle dechanneling lengths are relatively shorter than for positive particles, negative particle channeling studies are possible at high energy with new short crystal approaches using quasimosaic crystals or anticlastic bends. These studies are interesting for planar and axial cases for normal bending, volume reflection, and volume capture. Possible mid and long range applications include antiproton colliders.

Muon channeling appears not to be a significant study tool because the effects of nuclear interactions in channeling are much less important than dechanneling. On the other hand the possibility of higher energy muon colliders is now discussed for the far future. There collimation may once again become a challenge and channeling might have something to offer.

This paper has not reviewed volume reflection and capture or dechanneling in these processes. The recent H8/H4 studies have now produced interesting results[4] bearing on this area. Efficiencies up to 94 to 98% have been observed. These experimental results also agree with Monte Carlo simulations. These very interesting results open up important new possibilities for accelerator beam collimation in the TeV regime.